\documentclass[preprint,showpacs,showkeys,preprintnumbers,amsmath,amssymb]{revtex4}

\usepackage{graphicx}
\usepackage{dcolumn}
\usepackage{bm}

\begin{document}

\preprint{}

\title{Hilbert-Schmidt Separability Probabilities and Noninformativity of
Priors}

\author{Paul B. Slater}%
\email{slater@kitp.ucsb.edu}
\affiliation{%
ISBER, University of California, Santa Barbara, CA 93106\\
}%
\date{\today}

\begin{abstract}
The Horodecki family employed the Jaynes maximum-entropy principle, fitting
the mean ($b_{1}$) of the Bell-CHSH observable ($B$). This model was extended
by Rajagopal by incorporating the dispersion ($\sigma_{1}^2$) of the
observable, and by Canosa and Rossignoli, by generalizing the observable 
($B_{\alpha}$).
We further extend the Horodecki {\it one}-parameter model in {\it both} these
manners, obtaining a {\it three}-parameter ($b_{1},\sigma_{1}^2,\alpha$)
two-qubit model, for which we find a highly interesting/intricate 
continuum $(-\infty < \alpha < \infty)$ 
of Hilbert-Schmidt (HS) 
separability probabilities --- in which, the {\it golden ratio} is featured. 
Our model can be contrasted
with the three-parameter ($b_{q}, \sigma_{q}^2,q$) one of Abe and Rajagopal, which employs 
a $q$(Tsallis)-parameter rather than $\alpha$, and has  simply 
$q$-{\it invariant}
HS separability probabilities of $\frac{1}{2}$.
Our results emerge in a study initially focused on embedding certain 
information metrics over the two-level quantum systems into a $q$-framework. 
We find evidence that Srednicki's recently-stated biasedness criterion
for {\it noninformative} priors yields rankings of priors fully consistent
with an information-theoretic test of Clarke, previously applied to quantum systems by Slater.
\end{abstract}

\pacs{Valid PACS 02.50.Tt, 03.67.-a, 05.30.-d, 89.70.+c}
\keywords{separability probabilities, maximum entropy, Hilbert-Schmidt metric, Bures metric, escort distribution, density matrices, Husimi distribution, comparative noninformativity, Fisher information, $q$ order-parameter, 
golden ratio, 
nonextensitivity/Tsallis index, Bayes' Theorem, posteriors, priors, monotone metrics}

\maketitle
\section{Introduction}
  
Both Rajagopal \cite{Raj}, as well as Canosa and Rossignoli 
\cite{canosa} have extended a well-known
maximum-entropy model of the Horodecki family \cite{horodeckis} to
{\it two}-parameter models, but in different fashions. Rajagopal incorporated the {\it dispersion} ($\sigma_{1}^2$) of the 
Bell-CHSH observable ($B$) \cite{clauser}, 
the {\it mean} ($b_{1}$) of which is
already fitted in the Horodecki model, while Canosa and Rossignoli fitted
the mean of {\it generalized} Bell-CHSH observables 
($B_{\alpha}$).
We combine their two approaches into a {\it three}-parameter ($b_{1},\sigma_{1}^2,\alpha$)
model, for which we uncover a very interesting continuum ($-\infty < \alpha
<\infty$) of 
exact Hilbert-Schmidt separability probabilities (sec.~\ref{trivariate}, 
Fig.~\ref{fig:HSsepprob}, (\ref{HSsepprobcases})) --- in which, among 
other features, the 
{\it golden ratio} \cite{livio,markowsky} appears.

Our model can be interestingly contrasted with a 
three-parameter ($b_{q},\sigma_{q}^2,q$) 
one also  
of Abe and Rajagopal 
\cite{aberajagopal} (sec.~\ref{ARTQ}), which incorporates the
$q$-parameter (nonextensitity/Tsallis index/escort parameter), rather than
the $\alpha$-parameter of $B_{\alpha}$ (\ref{canosanew}).
The continuum (over $q$) 
of separability probabilities ({\it independently} of the metric
employed) is simply a constant (equal to $\frac{1}{2}$ in 
the Hilbert-Schmidt case, and to the ``silver mean'', 
$\sqrt{2}-1$, for the Bures and other monotone metrics). 
We examine a {\it two}-parameter
($b_{1}, \alpha$)
Canosa-Rossignoli-type model also exhibiting an interesting 
(non-flat) continuum 
(sec.~\ref{bivariatesec}, Fig.~\ref{fig:HSsepprob2}, (\ref{anotherinstance})).
Additionally, we 
obtain exact (Hilbert-Schmidt and Bures) separability probabilities
for the three-parameter Tsallis-Lloyd-Baranger model \cite{tlb} 
(sec.~\ref{TLBsec})

Our results emerge in a study initially focused on embedding certain 
information metrics over the two-level quantum systems into a 
$q$-framework. 
We find evidence (sec.~\ref{maincomp}, Fig.~\ref{fig:biasedness}) 
that Srednicki's recently-stated biasedness criterion
for {\it noninformative} priors \cite{srednicki} 
yields rankings of priors fully consistent
with an information-theoretic test of Clarke \cite{clarke}, previously applied to quantum systems by Slater \cite{compnoninform}.

\section{Noninformativity of Priors}

Some fifteen years ago, Wootters asserted that ``there does not seem to be any natural measure on the set of all mixed states'' \cite[p. 1375]{wootters}. He did, however, consider random density matrices with all eigenvalues {\it fixed}.
He remarked that once ``the eigenvalues are fixed, then all the density 
matrices in the ensemble are related to each other by the unitary group, so
it is natural to use the unique unitarily invariant measure to define the
ensemble'' \cite[p. 1375]{wootters} (cf. \cite{mjwhall}).

Arguing somewhat similarly,
Srednicki recently proposed that
in choosing a prior distribution over density matrices, ``we can use the
principle of indifference, applied to the unitary symmetry of Hilbert space,
to reduce the problem to one of choosing a probability distribution for the
eigenvalues of $\rho$. There is, however, no compelling rationale for any
particular choice; in particular, we must decide how biased we are towards
pure states'' \cite[p. 6]{srednicki}.

\begin{figure}
\includegraphics{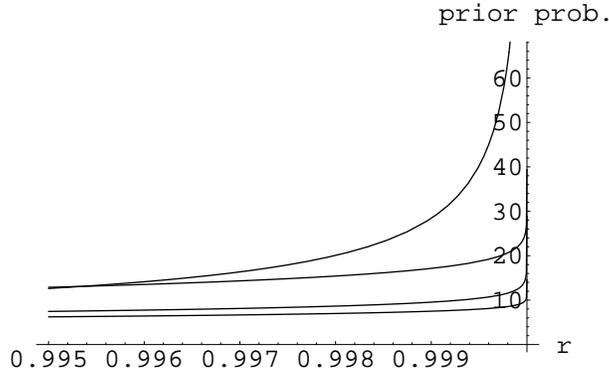}
\caption{\label{fig:biasedness}Four 
univariate marginal prior probability 
distributions in the near-to-pure-state region $r \in [1-\epsilon, 1]$, 
where $r$ is the radial coordinate in the Bloch sphere representation
of two-level quantum systems, and $r=1$ corresponds to a pure state. 
The order of dominance
{\it fully} complies with that (\ref{ordering}) obtained by
the information-theoretic-based 
comparative noninformativity test}
\end{figure}

To be specific, we find, 
in an analysis involving four prior probabilities ($p$'s),
that the {\it information-theoretic}-based comparative 
noninformativity test devised by Clarke yields a ranking 
\begin{equation} \label{ordering}
p_{F_{q=1}} > p_{B} > p_{B_{q=1}trunc} >p_{F}
\end{equation}
{\it fully} 
consistent  (Fig.~\ref{fig:biasedness}) with Srednicki's 
recently-stated criterion for priors
of ``biasedness to
pure states''. Two of the
priors are formed
by {\it extending} certain metrics of quantum-theoretic interest
from three- to four-dimensions --- by incorporating 
the $q$-parameter.
The 
three-dimensional metrics are the 
Bures (minimal monotone) metric over the two-level quantum systems and 
the Fisher information metric over the corresponding 
family of Husimi distributions.
The priors $p_{B}$ and $p_{F}$ are the ({\it independent}-of-$q$) 
normalized volume elements of these metrics, 
and $p_{F_{q=1}}$ is the normalized volume element of the $q$-{\it extended} 
Fisher information metric, with $q$ set to 1. 
While 
originally intended  to similarly be
the $q$-extension of the Bures metric, with
$q$ then set to 1, the prior $p_{B_{q=1}trunc}$, actually
entails 
the {\it truncation} of the only 
{\it off}-diagonal entry of the extended Bures 
metric
tensor. Without this truncation, 
the $q$-extended Bures volume element is  {\it null}, as is also the case 
in three {\it higher}-dimensional  
quantum scenarios we examine 
(including the Abe-Rajagopal two-qubit 
states (sec.~\ref{ARTQ}), for which we further find $q$-{\it invariant} 
Bures separability probabilities equal to $\sqrt{2}-1$ [the ``silver mean'']
and Hilbert-Schmidt ones equal to $\frac{1}{2}$ (sec.~\ref{qInvariance}), 
and the Tsallis-Lloyd-Baranger two-qubit states (sec.~\ref{TLBsec})).
\section{Bures Metric}
The Bures (minimal monotone) metric --- the volume element of which we 
normalize to obtain one ($p_{B}$) 
of the four prior probability distributions 
of 
principal
interest here --- yields the 
statistical distance between neighboring 
mixed quantum states ($\rho$) \cite{sam,uhlmann}.
It provides an embedding of 
the Fubini-Study metric \cite[sec. IV]{petzsudar}, which gives
the statistical distance between neighboring
pure quantum states ($|\psi \rangle$) (cf. \cite{majtey}).
H\"ubner gave an explicit formula for the Bures distance 
\cite[p. 240]{Hubner} (cf. \cite{luozhang}),
\begin{equation} \label{hub1}
d_{B}(\rho_{1},\rho_{2})^2 
= 2 -2 \mbox{tr} (\rho_{1}^{1/2} \rho_{2} \rho_{1}^{1/2})^{1/2}.
\end{equation}
Further, he expressed it 
in infinitesimal form as \cite[eq. (10)]{Hubner}
\begin{equation} \label{hub2}
d_{B}(\rho,\rho +d \rho)^2 =\Sigma_{ij} \frac{1}{2} \frac{|<i|d \rho| j>|^2}{\lambda_{i}+\lambda_{j}},
\end{equation}
where the $\lambda_{i}$'s are the eigenvalues  and the 
$|i \rangle$'s, the eigenvectors of $\rho$.
\subsection{Three-Dimensional Case} \label{sec3D}
In \cite{slaterBures}, using the familiar Bloch sphere (unit ball in Euclidean
3-space) representation of the two-level quantum systems ($2 \times 2$ 
density matrices),
\begin{equation} \label{nonescortDensityMatrix}
\rho = \frac{1}{2}  \left( \begin{array}{ccc}
1+z & x- i y \\
 x+ i y & 1-z\\
\end{array} \right), \hspace{.5in} r^2 = x^2+y^2+z^2 \leq 1,
\end{equation}
it was found (cf. \cite[p. 128]{mjwhall}), here converting from cartesian to spherical coordinates,
\begin{equation} \label{sphcoord}
x=r \cos{\theta_{1}}, \hspace{.3in} y=r \sin{\theta_{1}} \cos{\theta_{2}}, 
\hspace{.3in} z=r \sin{\theta_{1}}
\sin{\theta_{2}},
\end{equation}
that
\begin{equation} \label{BuresMetric}
d_{B}(\rho,\rho +d \rho)^2=\frac{1}{4} \Big(\frac{1}{(1-r^2)} dr^2 
+dn^2 \Big).
\end{equation}
The term $dr^2$ corresponds to the radial component of the metric
and $dn^2$, the tangential component ($dn^2= r^2 d \theta_{1}^2
+r^2 \sin^2{\theta_{2}}$).
In the setting of the quantum 
{\it monotone} metrics --- the 
Bures metric serving as the
{\it minimal} monotone one --- it is appropriate to express the tangential component of 
the Bures metric 
(\ref{BuresMetric}) in the form \cite[eq. (3.17)]{petzsudar},
\begin{equation} \label{unextendedBures}
\Big( (1+r) f_{B}(\frac{1-r}{1+r}) \Big)^{-1},
\end{equation}
where $f_{B}(t)= \frac{1+t}{2}$ is an {\it operator monotone} function 
\cite{lesniewski}.

The volume element of the Bures  metric 
(\ref{unextendedBures}) is $\frac{r^2 \sin{\theta_{1}}}{ 8 (1-r^2)}$,
which can be normalized to a {\it prior} probability distribution
over the Bloch sphere,
\begin{equation} \label{Buresprior}
p_{B}= \frac{r^2 \sin{\theta_{1}}}{\pi^2 (1-r^2)}.
\end{equation}
\subsection{Four-Dimensional Case}
Now, we can construct a {\it four}-dimensional family of 
(properly normalized/unit trace) $2 \times 2$ {\it escort} density
matrices (cf. \cite{naudts}),
\begin{equation} \label{escortDensityMatrix}
\rho_{\{q\}}= \Big( (1-r)^q+(1+r)^q \Big)^{-1}  \left( \begin{array}{ccc} 
1+z & x- i y \\ 
 x+ i y & 1-z\\
\end{array} \right)^q,
\end{equation}
for which $q=1$ recovers the standard Bloch sphere  representation 
(\ref{nonescortDensityMatrix}).
Applying H{\"u}bner's 
formula (\ref{hub2}), we have found that the {\it extended}
Bures metric (now incorporating the $q$-parameter)
has the form
\begin{equation} \label{extendedBures}
d_{Bures_q}(\rho,\rho+d \rho)^2= \frac{1}{4 (1+W^q)^2} \Big( 
W^q \log^2{W} dq^2 + \frac{4 q W^q \log{W}}{r^2-1} dq dr + 
\end{equation}
\begin{displaymath}
+ 4 \frac{q^2 W^q}{(r^2-1)^2 } 
dr^2
+ \frac{(-1+W^q)^2}{r^2 } dn^2 \Big),
\end{displaymath}
where $W=\frac{1-r}{1+r}$, that is, the ratio of the two 
eigenvalues of $\rho$.

\begin{figure}
\includegraphics{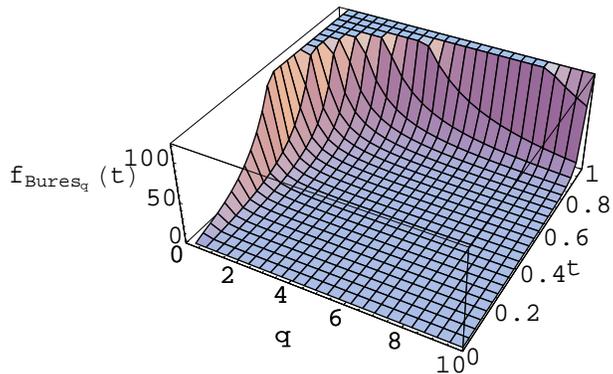}
\caption{\label{fig:monotonicgeneralization}The function $f_{Bures_{q}}(t)$
that yields the {\it tangential} component (\ref{generalf}) of the extended 
(four-dimensional) Bures metric (\ref{extendedBures})}
\end{figure}

The {\it tangential} component of the metric (\ref{extendedBures}) can
be expressed as $((1+r) f_{Bures_{q}}(W))^{-1}$, where
\begin{equation} \label{generalf}
f_{Bures_{q}}(t) =\frac{2 (1+t) (1+t^q)^2}{(-1+t^q)^2}.
\end{equation}
This 
bivariate function appears (Fig.~\ref{fig:monotonicgeneralization}) 
to be monotonically-increasing for any
fixed $q$ (cf. \cite{petzsudar}).

Now, in the earlier stage of our 
analyses, due to a programming oversight, we were under
the impression that the off-diagonal $dq dr$ term of (\ref{extendedBures})
was simply zero. If we do employ the fully correct form, 
with this $dq dr$ term included, we find that the
volume element is {\it null}. This, of course, could not yield a 
meaningful prior probability distribution. However, 
having proceeded under the impression that the $dq dr$ term was null,
we obtained a number of results that appear to be of interest 
and of some relevance.
Therefore, for much of this study, we will treat the $dq dr$ term as null, and
thus deal with a {\it truncated} $q$-extended Bures metric.

In the context of the harmonic oscillators states, 
Pennini and Plastino have argued that, in addition to the physical 
lower bound (ignorance-amount) of $q \geq 0$  that 
in a quantal regime,  
$q$ can be no 
{\it less}
than 1 \cite{pennini} --- due to the Lieb bound
on the Wehrl entropy \cite{lieb}. However, for the two-level quantum systems 
to the study of which we
restrict ourselves here, the lower  
bound on the Wehrl entropy is $\frac{1}{2}$  \cite[eq. (12)]{schupp}.
We, thus, consider $q \in [\frac{1}{2},\infty]$ to be the range of possible
values of the escort parameter $q$. In practice, though, we will, for numerical  and graphical 
purposes and normalization of the (divergent over $q \in [1/2,\infty]$) 
truncated extended Bures 
volume element (Sec.~\ref{qInference}), 
consider that $q \in [\frac{1}{2},500]$.

\begin{figure}
\includegraphics{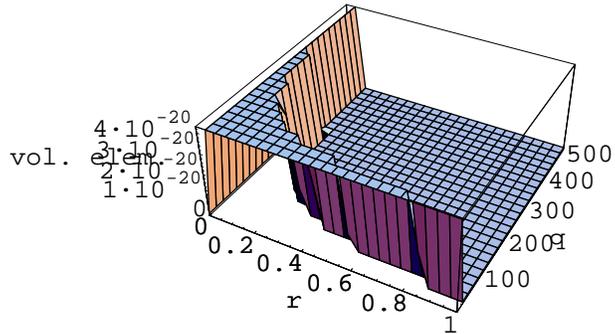}
\caption{\label{fig:BuresqVolElem}Two-dimensional marginal of the 
{\it truncated} four-dimensional extended Bures volume element (\ref{extendedBures})}
\end{figure}

\begin{figure}
\includegraphics{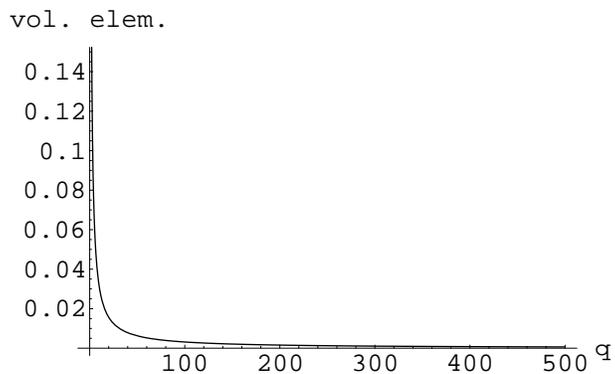}
\caption{\label{fig:BuresqVolElem1}One-dimensional marginal 
(\ref{exactprior}) over $q$ of the 
four-dimensional {\it truncated} 
extended Bures volume element (\ref{extendedBures})}
\end{figure}

In Fig.~\ref{fig:BuresqVolElem} we show the {\it two}-dimensional 
{\it marginal}
volume element of (\ref{extendedBures}) (after omission of the
$dq dr$ term)  --- integrating out the spherical angles, $\theta_{1}, \theta_{2}$, and leaving the radial coordinate $r$ and 
the escort parameter $q$.
In Fig.~\ref{fig:BuresqVolElem1}, further integrating out $r$, 
we show the corresponding {\it one}-dimensional marginal
volume element of (\ref{extendedBures}) (after omission of the $dq dr$ 
term) over $q$.
This (Fig.~\ref{fig:BuresqVolElem1}) has the exact expression 
\begin{equation} \label{exactprior}
\frac{\pi (1+\log{4})}{24 q}.
\end{equation}
This prior, thus, conforms to Jeffreys' rule --- as opposed to the
Bayes-Laplace rule, which would give a {\it constant} prior \cite{slaterlavenda}.

\begin{figure}
\includegraphics{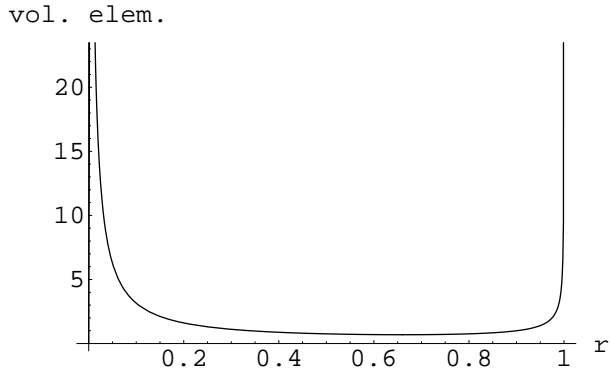}
\caption{\label{fig:BuresrVolElem1}One-dimensional marginal (obtained from 
(\ref{exactprior2})) over $r$ of the
four-dimensional extended Bures volume element (\ref{extendedBures}) after
omission of the off-diagonal $dq dr$ term}
\end{figure}

In Fig.~\ref{fig:BuresrVolElem1}, we integrate out $q \in [\frac{1}{2},
500]$,
leaving a (deep bowl-shaped)
 one-dimensional marginal over $r \in [0,1]$. (The corresponding 
marginal in the unextended Bures case is $\frac{\pi r^2}{2 (1-r^2)}$, 
so it is simply increasing with $r$, in that case.)
The associated  indefinite integral over $q$ is
\begin{equation} \label{exactprior2}
\frac{\pi \,\left( q\,W^q\,\left( 3 + W^{2\,q} \right) \,
       \log (W) - \left( 1 + W^q \right) \,
       \left( 2\,W^q + {\left( 1 + W^q \right) }^2\,
          \log (1 + W^q) \right)  \right) }{6\,
    \left( -1 + r^2 \right) \,
    {\left( 1 + W^q \right) }^3\,\log (W)}.
\end{equation}
(So, we obtain the function plotted in  
Fig.~\ref{fig:BuresrVolElem1} by substituting $q=500$ and $q=\frac{1}{2}$
into (\ref{exactprior2}) and taking the difference.)

For $q=1$, the {\it extended} Bures metric 
(\ref{extendedBures}) reduces to
\begin{equation} \label{extendedq1Bures}
ds_{Bures_{q=1}}(\rho,\rho+d \rho)^2= \frac{1}{16} (1-r^2) \log^2{W} dq^2 
- \frac{1}{4} \log{W} dq dr 
+ ds_{B}(\rho,\rho+d \rho)^2.
\end{equation}
Normalizing the volume element of this metric --- but 
first nullifying the off-diagonal $dq dr$ term --- to a (non-null) 
prior 
probability distribution over the Bloch
sphere, we obtain (cf. (\ref{Buresprior})),
\begin{equation}
p_{B_{q=1}trunc} =\frac{3}{4} \frac{r^2 \sin{\theta_{1}} \log{\frac{1}{W}}}{\pi (1 +\log{4})},
\end{equation}
one of the four priors that we rank (Fig.~\ref{fig:biasedness} and (\ref{ordering})) both by the comparative 
noninformativity test and Srednicki's biasedness criterion.

\subsection{Comparative Noninformativities in the Bures Setting} 
The {\it relative entropy} (Kullback-Leibler 
distance/information gain \cite{lisa,vedralRMP}) 
of $p_{B}$ with respect to 
$p_{B_{q=1}trunc}$ [which we 
denote $S_{KL}(p_{B},p_{B_{q=1}trunc})$] --- that is,
the {\it expected} value with respect to $p_{B}$
of $\log{\frac{p_{B}}{p_{B_{q=1}trunc}}}$ --- is
0.101846 ``nats'' of information. Now, reversing arguments, 
$S_{KL}(p_{B_{q=1}trunc},p_{B}) =0.0661775$. (We 
use the {\it natural} logarithm, and
not 2 as a
base, with one nat equalling 0.531 bits.)
Let us convert --- using Bayes' rule --- these 
two (prior) probability distributions to {\it posterior}
probability distributions ($post_{B}$ and $post_{Bures_{q=1}}$), 
by assuming {\it three}  
pairs of spin measurements, 
{\it one}  each
in the x-, y- and z-direction, each pair yielding one ``up'' and one ``down''.
This gives us the {\it likelihood} function (cf. \cite[eq. (9)]{srednicki} 
\cite[eq. (4.2)]{bagan}),
\begin{equation} \label{lik}
L(x,y,z)= \frac{(1-x^2) (1-y^2) (1-z^2)}{64}
\end{equation}
(which we convert to the spherical coordinates (\ref{sphcoord}) 
in which we perform our Mathematica computations).

Then, we have $S_{KL}(post_{B}||p_{B_{q=1}trunc})= 0.169782$ and 
$S_{KL}(post_{Bures_{q=1}}||p_{B}) = 0.197657$.
The relative magnitudes of the information gains obtained by passing from priors to 
posteriors (0.101846 to 0.169782 and 0.0661775 to 0.197657) seems to suggest 
that $p_{B}$ is  
somewhat {\it more} noniformative than
$p_{B_{q=1}trunc}$. This is 
confirmed, using the 
testing structure given in 
\cite{compnoninform,slaterHusimi} (cf. \cite{srednicki}),
if we {\it formally} use a likelihood ($L(x,y,z)^{\frac{1}{2}}$), 
which is the square root of (\ref{lik}), to compute $post_{B}$ and $post_{Bures_{q=1}}$.
Then, we see a {\it decrease} in relative entropy from 0.101846 to
0.093849 and an {\it increase} from 0.0661775 to 0.114669. So, $p_{B}$
can be made {\it closer} to $p_{B_{q=1}trunc}$ by {\it adding} information
to it, but not {\it vice versa}, leading us to conclude that
$p_{B}$ is {\it more} noninformative than $p_{B_{q=1}trunc}$, 
since it assumes {\it less} about the data. 
(Let us note, 
however, that in the class of monotone metrics \cite{petzsudar}, the 
Bures or minimal monotone metric appears to be the {\it least}
noninformative (cf. \cite[sec. 5]{mjwhall}). The {\it maximal} monotone metric, on the other 
hand,  is {\it not}
normalizable to a proper prior probability distribution over 
the Bloch sphere \cite{compnoninform}. So, there is an interesting question of 
whether there exists a 
{\it single}, distinguished {\it normalizable} monotone metric which is
{\it maximally} noninformative.)
\section{Fisher Information Metric of Husimi Distributions}
Let us now move to a classical context, employing the 
(generalized) Husimi distributions 
\cite{monge}, 
rather than density matrices to represent the two-level quantum 
systems. 
Use of the Fisher information 
(monotone) metric \cite{chentsov,pap} is 
now indicated. To generate the (properly normalized) 
{\it escort} Husimi distributions 
($H_{\{q\}}$) (cf. \cite{pennini}), from 
the Husimi distribution ($H = H_{\{1\}}$), we employ 
the formula (cf. (\ref{escortDensityMatrix})),
\begin{equation}
H_{\{q\}} =2 \left( r + q\,r \right) \Big( -{\left( 1 - r \right) }^{1 + q} +
    {\left( 1 + r \right) }^{1 + q} \Big)^{-1} H^{q}.
\end{equation}

The tangential components of the Fisher information metric 
for the escort Husimi distributions 
($H_{\{q\}}$) are of
the form \newline 
$((1+r) f_{F_{q}}(t))^{-1}$, where \cite[eq. (29)]{slaterHusimi}
\begin{equation} \label{nonintegral}
f_{F_{q}}(t) = \frac{\left( -1 + q \right) \,{\left( -1 + t \right) }^2\,
    \left( -1 + t^{1 + q} \right) }{q\,
    \left( 1 + t \right) \,
    \left( 1 - q + t + q\,t - t^q - q\,t^q - t^{1 + q} +
      q\,t^{1 + q} \right) }.
\end{equation}
In \cite[sec. V.D]{slaterHusimi}, 
we succeeded in finding similarly
general (for all $q$) 
formulas for the denominators, but not the numerators, 
of the radial components.

\begin{figure}
\includegraphics{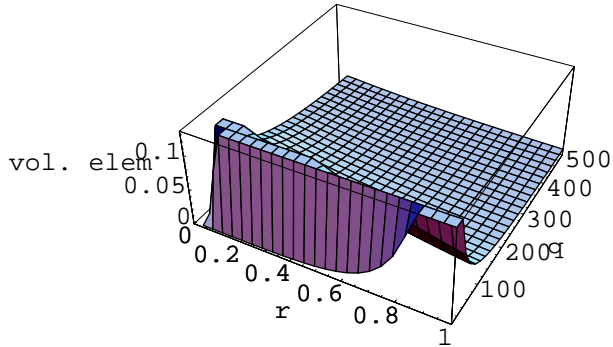}
\caption{\label{fig:HusimiqVolElem}Two-dimensional 
marginal of the four-dimensional extended 
Husimi volume element (\ref{extendedHusimi})}
\end{figure}

\begin{figure}
\includegraphics{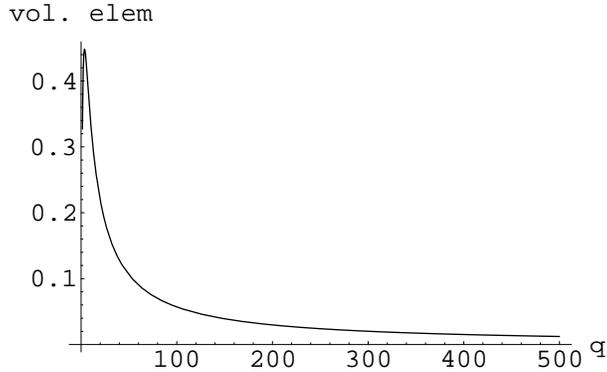}
\caption{\label{fig:HusimiqVolElem1}One-dimensional marginal over $q$ of the
four-dimensional extended Husimi volume element (\ref{extendedHusimi}). 
There is a peak near $q=3.59782$}
\end{figure}

\begin{figure}
\includegraphics{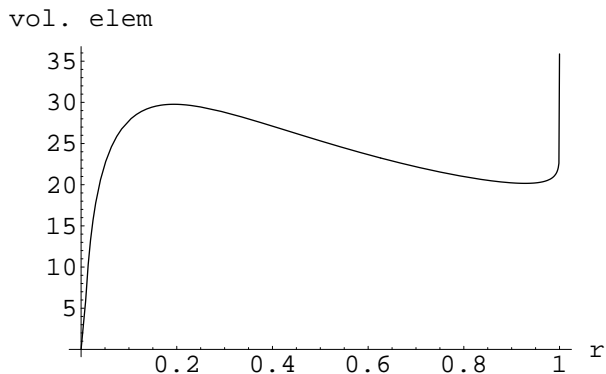}
\caption{\label{fig:HusimirVolElem1}One-dimensional marginal over $r$ of the
four-dimensional extended Husimi volume element (\ref{extendedHusimi}).
The upturn near $r=1$ may be due to (hard-to-avoid) numerical inaccuracy.}
\end{figure}

In Fig.~\ref{fig:HusimiqVolElem} we show (having to resort to some {\it 
numerical} 
integrations, since 
we lack explicit [$q$-general] expressions for certain of the metric 
elements) 
the counterpart to Fig.~\ref{fig:BuresqVolElem} 
for the four-dimensional extended {\it Husimi} metric.
Continuing with our numerical methods, we obtain the 
interesting unimodal curve (Fig.~\ref{fig:HusimiqVolElem1}) --- the peak being near $q=3.59782$, with a value there of 0.448488. This 
portrays the {\ one}-dimensional marginal Husimi volume element over $q$ (cf. Fig.~\ref{fig:BuresqVolElem1}).
In Fig.~\ref{fig:HusimirVolElem1} we show the 
(quite difficult-to-compute) one-dimensional marginal
over $r$ (cf. Fig.~\ref{fig:BuresrVolElem1}). (It appears the upturn 
near $r=1$ may be 
simply a numerical artifact. The difficulty consists in that, in some sense, we have to repeatedly perform  numerical integrations using results of other numerical integrations. It would be of interest to see how the curve changes as
the range of $q \in [\frac{1}{2},500]$ is modified.)
\subsection{Three-dimensional metric}

For the case $q=1$, the (unextended) three-dimensional 
Fisher information metric 
over the family of Husimi distributions takes the 
form \cite[eq. (2)]{slaterHusimi}
\begin{equation} \label{expressionHus}
ds_{F}(\rho,\rho+ d \rho)^2 = \frac{-2 r - \log (\frac{1-r}{1+r})}{2 r^3} dr^2 +
\Big((1+r) f_{F}(\frac{1-r}{1+r})\Big)^{-1} dn^2.
\end{equation}
Here,
\begin{equation} \label{zzz}
f_{F}(t)= \frac{(t-1)^3}{t^2-2 t \log{t}-1},
\end{equation}
which is the limiting case ($q \to 1$) of (\ref{generalf}).
To normalize the  volume element of this metric 
(\ref{expressionHus}) to a prior probability
distribution ($p_{F}$), we divide it by 1.39350989 \cite{slaterHusimi}.
\subsection{Four-dimensional metric}
In the extended (four-dimensional) 
case (cf. (\ref{extendedq1Bures})), {\it after} having set $q=1$,
we have,
\begin{equation} \label{extendedHusimi}
ds_{F_{q=1}}(\rho,\rho+ d \rho)^2= 
\Big( \frac{1}{4} -\frac{(-1+r^2)^2 \log^2{W}}{16 r^2} \Big) 
dq^2 
\end{equation}
\begin{displaymath}
+ \frac{2 r - (-1+r^2) \log{W}}{2 r^2} dq dr +ds_{F}(\rho,\rho+d \rho)^2.
\end{displaymath}
(So, the metric tensor here, in the same manner as in
the {\it untruncated} extended Bures case (\ref{extendedBures}), is not fully 
diagonal. We do {\it not} truncate the $q$-extended Fisher information
metric (\ref{extendedHusimi}) in any of our analyses.)
To normalize its (non-null) volume element 
to a prior probability distribution 
($p_{F_{q=1}}$) over the Bloch sphere, we must divide by 
0.24559293.
\section{Comparative Noninformativity Analysis} \label{maincomp}
We have that 
$S_{KL}(p_{F}||p_{F_{q=1}}) = 0.229666$
and $S_{KL}(p_{F_{q=1}}||p_{F}) = 0.170145$.
Further, using the likelihood (\ref{lik}), based on six hypothetical
measurements to generate posteriors, 
we obtain $S_{KL}(post_{F},p_{F_{q=1}}) = 0.70766$ and
$S_{KL}(post_{F_{q=1}}||p_{F}) =0.0641738$.
So, the comparative noninformativity test, which was initially developed by
Clarke \cite{clarke}, leads us to a firm conclusion that 
the four-dimensional-based probability distribution $p_{F_{q=1}}$
is {\it more} noninformative in nature than the three-dimensional-based
$p_{F}$.

Additionally, $S_{KL}(p_{B}||p_{F_{q=1}})=0.148269$ and 
$S_{KL}(p_{F_{q=1}}||p_{B}) = 0.0989669$. These are converted, 
respectively, to 0.283218 and 0.0842879 if we replace  the first arguments
of the two relative entropy functionals by posterior distributions based on
the (formal) square root ($L(x,y,z)^{\frac{1}{2}}$) of the likelihood function (\ref{lik}).
Thus, we can conclude that $p_{F_{q=1}}$ is also 
{\it more}
noninformative than $p_{B}$.

Further, $S_{KL}(p_{B_{q=1}trunc}||p_{F_{q=1}})=0.105463$
and $S_{KL}(p_{F_{q=1}}||p_{B_{q=1}trunc})=0.0914175$. 
Again, using the formal square root ($L(x,y,z)^{\frac{1}{2}}$) 
of the likelihood, we obtain
changes, respectively, to 0.245602 and 0.0408236. 
So, our
conclusion here is that $p_{F_{q=1}}$ is also more noninformative
than $p_{B_{q=1}trunc}$.
We already know from \cite{slaterHusimi} that $p_{B}$ 
is considerably more noninformative than $p_{F}$.

Continuing along these lines, 
$S_{KL}(p_{B_{q=1}trunc}||p_{F})=0.0191948$ and 
$S_{KL}(p_{F}||p_{B_{q=1}trunc})= 0.0234599$ (so the two 
distributions are relatively
close to one another).
Using  ($L(x,y,z)^{\frac{1}{2}}$)
to generate posterior distributions, the first statistic is altered 
(slightly decreased)
to 0.0143147, while the second statistic jumps to 0.1047772.

So, assembling these several relative entropy statistics,
we have the previously indicated ordering of the four priors (\ref{ordering}).
(The conclusions of the comparative noninformativity test appear to be
{\it transitive} in nature, although I can cite no explicit theorem to that 
effect.)
\subsection{Relation to Srednicki's Criterion for Priors}
In Fig.~\ref{fig:biasedness}, we show the one-dimensional
marginal probabilities of the four prior probabilities  over the radial 
coordinate $r$ in the near-to-pure-state range $r \in [.995,1]$.
The dominance ordering in this plot {\it fully} complies with
that (\ref{ordering}) found by the 
information-theoretic-based comparative noninformativity test. 
(We note that this ordering is {\it not} simply {\it reversed} near to the 
fully mixed state [$r=0$].)
Conjecturally,  
this could be seen as a specific case of some (yet unproven)
theorem --- perhaps utilizing the convexity and 
decreasing-under-positive-mappings 
properties \cite[p. 35]{ohya} of the relative entropy functional.

 So, the  information-theoretic (comparative-noninformativity)
test appears to incorporate Srednicki's criterion of
``biasedness to pure states'' \cite{srednicki}.
(Of course, it would be interesting to test the consistency between the 
comparative noninformativity test and Srednicki's criterion with a larger 
number of priors, as well as in higher-dimensional quantum settings 
(cf. \cite{slaterSPIN}).)
Srednicki does not explicitly 
observe that increasing biasedness to pure 
states corresponds to increasing noninformativity. He asserts
that ``we must decide how biased we are towards pure states''.

Srednicki focused on {\it two} possible priors. One was the uniform distribution over the Bloch sphere (unit ball). In \cite[sec. 2.2]{compnoninform},
we had concluded that this distribution was {\it less} noninformative than
$p_{B}$, in full agreement with contemporaneous work of 
Hall \cite{mjwhall}.
The second prior (``the Feynman measure''), which Srednicki points out is less biased to the pure states than the uniform distribution, 
was discussed in \cite{slaterlmp}. Neither of the two priors analyzed by
Srednicki corresponds to the normalized volume element of a {\it monotone} 
metric 
\cite{compnoninform,slaterlmp}.
\section{$q$-Extended Inference} \label{qInference}
In the setting of the $q$-parameterized escort density matrices
(\ref{escortDensityMatrix}),
the factor $\frac{1-z^2}{4}$ in the likelihood (\ref{lik}), giving the probability
(in the standard three-dimensional Bloch sphere setting) 
of one spin-up and one spin-down being measured 
in the $z$-direction, would be 
{\it replaced} by 
\begin{equation} \label{extendedlikelihood}
L_{q}(z) = \frac{r^2 (1+W^q)^2 -(-1+W^q)^2 z^2}{4 r^2 (1+W^q)^2},
\end{equation}
and similarly for the $x$- and $y$-directions. 
(For $q=1$, we recover $\frac{1-z^2}{4}$.)

It would be interesting to ascertain if the 
volume elements of the extended four-dimensional 
(truncated) Bures 
and
Husimi metrics ((\ref{extendedBures}) and (\ref{extendedHusimi})) could be integrated over the product of the Bloch sphere
{\it and} $q \in [\frac{1}{2},\infty]$ and normalized to (prior) probability distributions.
Then, using likelihoods 
incorporating the form (\ref{extendedlikelihood}), one 
could conduct the comparative noninformativity test in a {\it four}-dimensional
setting, rather than only the {\it three}-dimensional one employed 
throughout this study.
It turns out, however, that the three-fold 
integral --- holding $q$ fixed --- of the truncated 
volume element of 
(\ref{extendedBures}) over the
Bloch sphere is given by our formula (\ref{exactprior}).
Therefore, the four-fold integral of the one-dimensional marginal 
over the indicated product region with 
$q \in [\frac{1}{2},\infty]$ must
{\it diverge}. So, to achieve a {\it proper} probability distribution one
would have to truncate $q$ above a certain value.

Continuing along these lines, we omitted 
$q$ above 500 (and below $q=\frac{1}{2}$) 
and normalized 
the volume element of the 
(truncated) 
extended Bures metric to a proper probability
distribution. Then, the information gain 
with respect to such a prior, using $L_{q}(z)$, 
is
0.0597923  nats of information, while a
 {\it single} up or down measurement yields 0.134651 nats, and  two
measurements along the same axis 
giving the same outcome leads to an information gain of
0.349601. The analogous three (slightly {\it larger}) 
statistics, working in the unextended
framework (where $q$ does not explicitly enter, and is implicitly
understood to equal 1), using $p_{B}$ as prior, 
are, respectively,  $\frac{7}{6}-\log{3} \approx 0.0680544$, and
\begin{equation}
\frac{8\, _{p}F_{q}(\{ \frac{1}{2},1,
         2\} ,\{ \frac{3}{2},\frac{5}{2}\} ,1) -
      \pi \,\left( -5 + \log (64) \right) -6 -12\, K   }{6\,
    \pi } \approx 0.140186,
\end{equation}
(where $_{p}F_{q}$ denotes a generalized hypergeometric function and 
 $K \approx 0.915965594177$ is Catalan's constant) 
and $ \frac{59}{30} -\log{5} \approx 0.357229$. (We encountered 
numerical difficulties using Mathematica in attempting to extend these analyses
to measurements conducted in more than one direction, unless we restricted
$q$ to a range no larger than on the order of 10.)

One might also consider 
the possible relevance of 
$q$-analogs of the Clarke comparative noninformativity 
test, using $q$-relative entropy (Kullback-Leibler) divergence
\cite{johal,hirokisuyari}.
\section{$q$-Extended Bures Metric for Higher-Dimensional Quantum Scenarios}
\subsection{Four-Variable $3 \times 3$ Density Matrices}
In \cite{slaterSPIN}, we considered an extension of the 
$2 \times 2$ density matrices (\ref{nonescortDensityMatrix}) to the 
$3 \times 3$ form (by incorporating an additional parameter $v$)
\begin{equation} \label{threebythreeDensityMatrix}
\rho = \frac{1}{2}  \left( \begin{array}{ccc}
v+z & 0 &  x- i y \\
0 & 2 -2 v & 0 \\
 x+ i y & 0 & v-z\\
\end{array} \right), \hspace{.5in} r^2 = x^2+y^2+z^2  \leq v^2; 
\hspace{.3in} 0 \leq v \leq 1,
\end{equation}
The Bures metric was found there to take the form
\begin{equation} \label{spin1extension}
d_{B_{n=3}}(\rho,\rho+ d \rho)^2 = 
\frac{1}{4} \Big( \frac{r^2-v}{ (1-v) (r^2-v^2)} dv^2 
+\frac{r}{r^2-v^2} dv dr +
+\frac{v}{v^2-r^2} dr^2 +\frac{1}{v} dn^2 \Big).
\end{equation}
(So, the tangential component is {\it independent} of $r$, as with 
(\ref{BuresMetric}) (cf. \cite{mjwhall}).)
Normalizing the volume element of (\ref{spin1extension}), 
we obtain the prior probability distribution 
\cite[eq. (18)]{slaterSPIN}
\begin{equation}
p_{B_{n=3}}
= \frac{3 r^2 \sin{\theta_{1}}}{4 \pi^2 v \sqrt{1-v} \sqrt{v^2-r^2}}.
\end{equation}
We have calculated that the (five-dimensional) $q$-extension of this metric
has a tangential component of the form
\begin{equation}
\frac{{\left( {\left( -r + v \right) }^q - 
       {\left( r + v \right) }^q \right) }^2}{4\,r^2\,
    \left( {\left( -r + v \right) }^q + 
      {\left( r + v \right) }^q \right) \,
    \left( {\left( 2 - 2\,v \right) }^q + 
      {\left( -r + v \right) }^q + 
      {\left( r + v \right) }^q \right) },
\end{equation}
but have not yet been able to derive simple forms for the other entries
of this metric tensor.

Numerical tests appear to indicate that the volume element
of this $q$-extended Bures metric tensor is (also) identically zero.
\subsection{Abe-Rajagopal  Two-Qubit States} \label{ARTQ}

Since our first two attempts above  to extend the Bures metric from an 
$n$-dimensional 
setting to an $(n+1)$-dimensional 
framework, by embedding the $q$ order parameter,
have 
yielded metrics (one of them being (\ref{extendedBures})) 
with zero volume elements, we were curious as to
whether or not we could obtain, in some other quantum context, a {\it 
nondegenerate}
$q$-extension of the Bures metric. In this regard, we 
turned our attention to the paper,
``Quantum entanglement inferred by the principle of maximum nonadditive
entropy'' of Abe and Rajagopal \cite{aberajagopal} (cf. \cite[eq. (14)]{tlb}).

Their principal object of study is a $4 \times 4$ density matrix
\cite[eq. (32)]{aberajagopal}, being ostensibly parameterized by
{\it three} variables, the order (nonadditivity) parameter $q$, 
the $q$-expected value $b_{q}$ of the Bell-CHSH observable and 
its dispersion $\sigma_{q}^2$. ({\it Two} of the four eigenvalues 
of the density matrix are always equal.)

We applied the H{\"u}bner formula (\ref{hub2}) for the Bures metric
to this family of $4 \times 4$ density matrices, considering $q$
as a freely-varying parameter, along with $b_{q}$ and $\sigma_{q}^2$.
Computing the $3 \times 3$ 
Bures metric tensor, and {\it then} setting 
$q=1$, we obtain the metric
\begin{equation} \label{AbeRajextended}
ds_{AbeRaj_{q=1}}(\rho,\rho+ d\rho)^2  = \frac{c}{1024} dq^2 +
\end{equation}
\begin{displaymath}
\frac{\log (-2\,{\sqrt{2}}\,{b_q} + {{{\sigma }_q}}^2) - 
    \log (2\,{\sqrt{2}}\,{b_q} + {{{\sigma }_q}}^2)}{8\,
    {\sqrt{2}}} dq db_{q} + 
\end{displaymath}
\begin{displaymath}\frac{2\,\log (8 - {{{\sigma }_q}}^2) - 
    \log (-2\,{\sqrt{2}}\,{b_q} + {{{\sigma }_q}}^2) - 
    \log (2\,{\sqrt{2}}\,{b_q} + {{{\sigma }_q}}^2)}{32} dq d\sigma_{q}^2 +
\end{displaymath}
\begin{displaymath}
\frac{\sigma_{q}^2}{-32 b_{q}^2 +4 (\sigma_{q}^2)^2} (db_{q})^2 +\frac{b_{q}}{16 b_{q}^2 -2 (\sigma_{q}^2)^2 } d b_{q} d \sigma_{q}^2 +\frac{b_{q}^2 -\sigma_{q}^2}{4 (-8 +\sigma_{q}^2) (-8 b_{q}^2 +(\sigma_{q}^2)^2)} (d \sigma_{q}^2)^2.
\end{displaymath}
Here, 
we have 
\begin{equation}
c = -4\,{\log (8 - {{{\sigma }_q}}^2)}^2\,{{{\sigma }_q}}^2\,
   \left( -8 + {{{\sigma }_q}}^2 \right)  + 
  2\,\log (-2\,{\sqrt{2}}\,{b_q} + {{{\sigma }_q}}^2)\,
   \log (2\,{\sqrt{2}}\,{b_q} + {{{\sigma }_q}}^2)\,
   \left( 8\,{{b_q}}^2 - 
     {{{{\sigma }}_q}}^4 \right)
\end{equation}
\begin{displaymath}
  - 
  {\log (-2\,{\sqrt{2}}\,{b_q} + {{{\sigma }_q}}^2)}^2\,
   \left( 8\,{{b_q}}^2 + 
     {{{\sigma }_q}}^2\,
      \left( -16 + {{{\sigma }_q}}^2 \right)  - 
     4\,{\sqrt{2}}\,{b_q}\,
      \left( -8 + {{{\sigma }_q}}^2 \right)  \right) 
\end{displaymath}
\begin{displaymath}
 - 
  {\log (2\,{\sqrt{2}}\,{b_q} + {{{\sigma }_q}}^2)}^2\,
   \left( 8\,{{b_q}}^2 + 
     {{{\sigma }_q}}^2\,
      \left( -16 + {{{\sigma }_q}}^2 \right)  + 
     4\,{\sqrt{2}}\,{b_q}\,
      \left( -8 + {{{\sigma }_q}}^2 \right)  \right) +
\end{displaymath}
\begin{displaymath}
  4\,\log (8 - {{{\sigma }_q}}^2)\,
   \left( -8 + {{{{\sigma }}_q}}^2 \right) \,
   \left( \log (-2\,{\sqrt{2}}\,{b_q} + 
        {{{\sigma }_q}}^2)\,
      \left( -2\,{\sqrt{2}}\,{b_q} + 
        {{{{\sigma }}_q}}^2 \right)  + 
     \log (2\,{\sqrt{2}}\,{b_q} + {{{\sigma }_q}}^2)\,
      \left( 2\,{\sqrt{2}}\,{b_q} + 
        {{{{\sigma }}_q}}^2 \right)  \right).
\end{displaymath}
Numerical computations indicate that the volume element of the metric 
$ds_{AbeRaj_{q}}(\rho,\rho+d \rho)^2$, for any value of $q$, is zero.

In the {\it unextended} (two-parameter) case, 
the {\it nondegenerate} volume element (with $q=1$) is 
\begin{equation} \label{specialcase}
dV_{AbeRaj_{q=1}}= \frac{{\sqrt{-\left( \frac{1}
         {\left( -8 + \sigma_{q} \right) \,
           \left( -8\,b_{q}^2 + (\sigma_{q}^2)^2 \right) } \right) }}}{4} 
d b_{q} d \sigma_{q}^2.
\end{equation}

\subsubsection{$q$-{\it Invariance} of Bures Volumes of 
Separable and Separable and Nonseparable AR States} \label{qInvariance}
\begin{figure}
\includegraphics{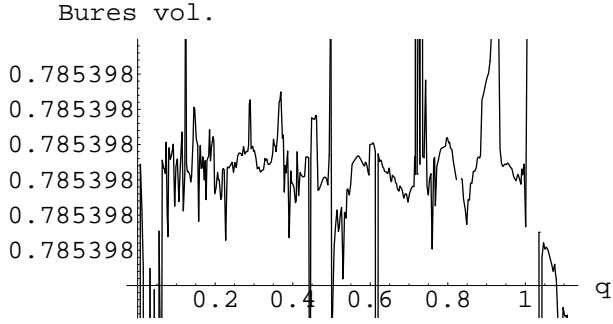}
\caption{\label{fig:ConstantBuresVol}Numerical integration estimates of the 
Bures volume of the (separable and
nonseparable) AR two-qubit 
states, as a function of $q$. This volume is 
{\it known} to be  $\frac{\pi}{4} \approx  0.7853981634$ for $q= \frac{1}{2}$ and 1 and appears to be so for all (positive) $q$.}
\end{figure}

\begin{figure}
\includegraphics{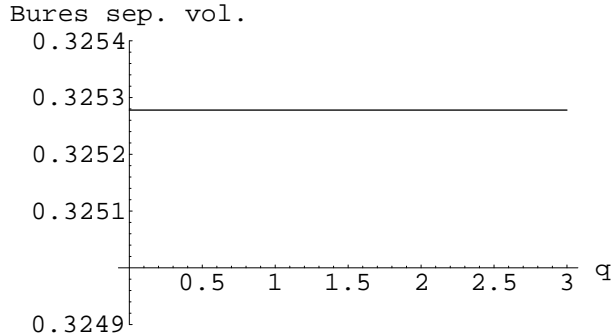}
\caption{\label{fig:ConstantBuresSepVol}Numerical integration 
estimates of the Bures volume of the separable 
A two-qubit 
states, as a function of $q$. This volume is {\it known} 
to be exactly $\frac{\pi (\sqrt{2}-1)}{4} \approx  0.325323$ for $q = \frac{1}{2}$ and 1 and appears to be so for all (positive) $q$.}
\end{figure}

In \cite{slaterEuro}, it was asserted that
for the cases $q=\frac{1}{2}$ and 1, the associated {\it separability
probabilities} of the Abe-Rajagopal (AR) states 
were equal to the ``silver mean'', that is, 
$\sqrt{2}-1 \approx 0.414214$ 
(cf. \cite{slaterJGP,slaterPRA}. We have reconfirmed these two {\it probabilities}, while also finding that the Bures
volume of separable and nonseparable states is, in both these cases, equal
to $\frac{\pi}{4} \approx 0.7853981634$. This also appears to be the Bures volume for all positive $q$, as indicated by the results obtained by numerical 
integration presented in Fig.~\ref{fig:ConstantBuresVol}.
The integrand employed (that is, the Bures volume element) was
\begin{equation}
dV_{AbeRaj}=  16 \sqrt{\frac{\left(8-\sigma _q\right)^{\frac{1}{q}-2}
   \left(\sigma _q-2 \sqrt{2} b_q\right)^{\frac{1}{q}}
   \left(2 \sqrt{2} b_q+\sigma
   _q\right)^{\frac{1}{q}}}{q^4 \left(\sigma _q^2-8
   b_q^2\right)^2 \left(2 \left(8-\sigma
   _q\right)^{\frac{1}{q}}+\left(\sigma _q-2 \sqrt{2}
   b_q\right)^{\frac{1}{q}}+\left(2 \sqrt{2} b_q+\sigma
   _q\right)^{\frac{1}{q}}\right)^3}} d b_{q} d \sigma_{q}^2.
\end{equation}
It also appears (Fig.~\ref{fig:ConstantBuresSepVol}) 
that the Bures volume of the separable (only) AR-states 
is equal to $\frac{\pi (\sqrt{2}-1)}{4} \approx 0.325323$ for {\it 
all}
positive $q$
and, thus, the separability probabilities (obtained by taking the ratios) 
are all 
simply $\sqrt{2}-1 \approx 0.414214$ (that is, the ``silver mean'').
(The numerical integration employed to generate Fig.~\ref{fig:ConstantBuresSepVol} is more challenging --- due to the necessary imposition of the 
Peres separability criterion --- than 
to create Fig.~\ref{fig:ConstantBuresVol}, so
we could not obtain as many significant digits.)

These $q$-invariance results stand in interesting contrast 
to the emphasis of Abe and Rajagopal ``that for $q>1$, indicating the subadditive feature of the Tsallis entropy, the entangled region is small and enlarges as one goes into the superadditive regime where $q<1$'' \cite[p. 3464 and Fig.~1]{aberajagopal}. 
But, in terms of the Bures metric (and others we will see below) the
{\it measure} of the region does {\it not} change with $q$.

Using the {\it Hilbert-Schmidt} metric 
\cite{hilb2}, rather than the Bures,
we find that the volume of separable and nonseparable AR
two-qubit states is {\it equal} to $\frac{1}{4 \sqrt{2}}$ for both
$q=\frac{1}{2}$ and 1 and the volume of separable states is equal to
$\frac{1}{8 \sqrt{2}}$ for both these values of $q$, 
so the corresponding 
Hilbert-Schmidt separability probabilities are simply $\frac{1}{2}$. 
(If we employ either the {\it Wigner-Yanase} [monotone] metric \cite{paolo} 
or the {\it arithmetic  average} [monotone] metric \cite{slaterPRA}, 
then, for $q=1$, we obtain
exactly the same [volume] results as using the Bures metric, and for $q=\frac{1}{2}$ --- using numerical rather than symbolic 
methods in the Wigner-Yanase case --- quite clearly  the same also.)
So, it certainly appears that the $q$-invariance of the total
and separable volumes of the AR-states is 
metric-{\it independent}. Canosa and Rossignoli \cite[p. 4]{canosa2} have noted that for the AR-states, the ``final maximum entropy density
is actually {\it independent of the choice of $f$}'', where $f$ is a smooth
{\it concave} function.

The Bures separability probability 
(as well as that based on the Wigner-Yanase metric) 
of the (one-parameter) ``Jaynes state'' \cite{horodeckis,Raj},
in which (unlike the AR-states) 
no constraint on the dispersion is present (and $q$ is implicitly equal to 1) 
, is
$\frac{2\,\arcsin ({\sqrt{2}} - 1)}{\pi } \approx 0.271887$.
(Again, note the presence of the silver mean --- and implicitly in the very 
next formula.) The Hilbert-Schmidt separability
probability is 
\begin{equation}
\frac{\sinh
   ^{-1}\left(2-\sqrt{2}\right)+\text{Root}\left[\text{$\#
   $1}^4-148 \text{$\#$1}^2+68\&,3\right]}{\sqrt{6}+\sinh
   ^{-1}\left(\sqrt{2}\right)} \approx 0.343602.
\end{equation}

To convert from the AR two-qubit density matrix for $q=1$ to that
for $q=\frac{1}{2}$, we merely have to perform the transformation
\begin{equation}
\left\{\sigma _1^2\to \frac{4 \left(8
   b_{\frac{1}{2}}^2+{(\sigma^2 _{\frac{1}{2}})}^2\right)}{4
   b_{\frac{1}{2}}^2+{(\sigma^2 _{\frac{1}{2}})}^2-8 {\sigma^2
   _{\frac{1}{2}}}+32},b_1\to \frac{8 b_{\frac{1}{2}}
   {\sigma^2 _{\frac{1}{2}}}}{4 b_{\frac{1}{2}}^2+{(\sigma^2
   _{\frac{1}{2}})}^2-8 {\sigma^2 _{\frac{1}{2}}}+32}\right\}.
\end{equation}
Presumably, there is a (more complicated, in general) transformation between
AR-states for any pair of distinct values of $q$.
So, in retrospect, the $q$-invariance of the (Bures, Hilbert-Schmidt, Wigner-Yanase and arithmetic average) metric volumes is not
so surprising, since we are simply working within one family of 
{\it two}-parameter 
density matrices, 
the various $q$-manifestations of which can be obtained by suitable
reparameterizations. Similarly, the null nature of the $q$-extended Bures metric for the AR-states can be seen in this light.

\subsubsection{Trivariate Jaynes state using generalized Bell-CHSH observables} \label{trivariate}
It would be interesting to extend and analyze the AR-states
based on modifications of
the Bell-CHSH observable (cf. \cite[sec. 3]{canosa} \cite{bcpp}).
In fact, we pursued such a line of investigation, using
\begin{equation} \label{canosanew}
B_{\alpha}= 2 \sqrt{2} (|\Phi^{+} \rangle \langle \Phi^{+} | - \alpha
|\Psi^{-} \rangle \langle \Psi^{-}|),
\end{equation}
as the observable, where for $\alpha=1$, we recover the Bell-CHSH observable
employed by Abe and Rajagopal \cite[eq. (6)]{aberajagopal} (cf. \cite[eq. (15)]{canosa}). (We utilized the Jaynes maximum entropy strategy \cite{horodeckis}, 
implicitly taking $q=1$ --- so, most precisely, we are extending the model
discussed by Rajagopal in \cite{Raj} to incorporate generalized Bell-CHSH observables or, alternatively, the Canosa-Rossignoli model to included the 
dispersion.) Then, the volume element of the Bures metric, considering
$\alpha$ as a parameter, in addition to the expectation $b_{1}$ and the
dispersion $\sigma_{1}$, was {\it null}. Considering, on the other hand,
$\alpha$ to be simply a fixed constant, the 
bivariate Bures volume element was
of the (non-null) 
form $\frac{1}{4} \sqrt{\frac{1}{C + D}} d b_{1} d \sigma_{1}^2$, where 
\begin{equation}
C = \left(8 \alpha -\sigma
   _1^2\right) (\sigma _1^2)^2 -4 \sqrt{2} (\alpha -1) b_1
   \left(\sigma _1^2-4 \alpha \right) \sigma _1^2 +
\end{equation}
and 
\begin{displaymath}
D = 16
   \sqrt{2} (\alpha -1) \alpha  b_1^3-8 b_1^2
   \left(\left(\sigma _1^2+8\right) \alpha ^2-3 \sigma
   _1^2 \alpha +\sigma _1^2\right).
\end{displaymath}
For $\alpha=1$, we recover (\ref{specialcase}), so the associated 
Bures separability probability is the silver mean. For $\alpha=1$, using the
HS-metric now, the total volume of states is $\frac{1}{4 \sqrt{2}}$ and that of
the separable states is $\frac{1}{8 \sqrt{2}}$, so 
the HS separability probability is simply $\frac{1}{2}$.

For the case $\alpha=2$, we obtained a result of 0.35368 for the Bures volume
of separable and nonseparable states, and 0.2000322 for the Bures volume of 
only separable states, yielding a separability probability of 0.566392.
The comparable results for the Hilbert-Schmidt case 
(the volume element --- {\it independent} of $b_{q}$ and 
$\sigma_{q}^2$ --- being $\frac{1}{32 \alpha (1+\alpha)} db_{q} 
d \sigma_{q}^2$) were $\frac{1}{12 \sqrt{2}}$
and $\frac{5}{96 \sqrt{2}}$, with a separability probability of $\frac{5}{8}$. 

Exact integration, then, gave the HS volume of separable and nonseparable
states to equal, {\it in general}, $\frac{1}{2 \sqrt{2} (\alpha +\alpha^2)}$, 
and that of the separable states --- but only for $\alpha  \geq 1$ --- to be $\frac{3 \alpha -1}{8 \sqrt{2} \alpha^2 (1+\alpha)}$, so the Hilbert-Schmidt separability probability for $\alpha \geq 1$ is
simply equal to $\frac{3}{4} -\frac{1}{4 \alpha}$. 
For $\alpha=\frac{1}{2}$, the HS separability probability appeared to be 
$\frac{37}{64} \approx 0.578125$.

\begin{figure}
\includegraphics{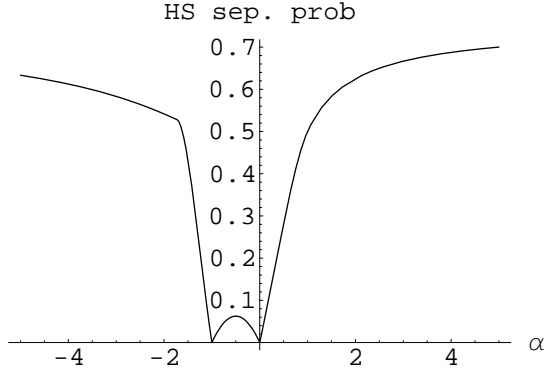}
\caption{\label{fig:HSsepprob}Separability probabilities (\ref{HSsepprobcases}) based on the Hilbert-Schmidt metric, as a function of the parameter $\alpha$, 
for the new class of trivariate Jaynes states employing
{\it generalized} Bell, Clauser, Horne, Shimony, Holt (Bell-CHSH) observables 
(\ref{canosanew})}
\end{figure}

Then, using the integration over implicitly defined regions 
feature new to Mathematica 5.1, we were able to obtain
the HS {\it separable} volumes, for all (real) values of $\alpha$,
\begin{equation}
\left(
\begin{array}{ll}
 -\frac{1}{8 \sqrt{2}} & -1<\alpha <0 \\
 \frac{5}{16 \sqrt{2}} & \alpha =-\sqrt{2} \\
 \frac{1}{32} \left(5 \sqrt{2}-\sqrt{10}\right) & \alpha
   =-\frac{1}{2}+\frac{\sqrt{5}}{2} \\
 \frac{1}{16} \left(2 \sqrt{2}-\sqrt{2} \alpha \right) &
   0<\alpha <-\frac{1}{2}+\frac{\sqrt{5}}{2}\lor
   -\frac{1}{2}+\frac{\sqrt{5}}{2}<\alpha <1 \\
 \frac{3 \sqrt{2} \alpha -\sqrt{2}}{16 \alpha ^2 (\alpha
   +1)} & \alpha \geq 1 \\
 \frac{3 \sqrt{2} \alpha ^2+\sqrt{2}}{16 \alpha ^2
   \left(\alpha ^2-1\right)} & \alpha \leq -\sqrt{3} \\
 \frac{-\sqrt{2} \alpha ^4+5 \sqrt{2} \alpha
   ^2-\sqrt{2}}{16 \alpha ^2} & -\sqrt{2}<\alpha <-1\lor
   -\sqrt{3}<\alpha <-\sqrt{2},
\end{array}
\right)
\end{equation}
and, dividing by the total HS volume  ($\frac{1}{2 \sqrt{2} 
(\alpha +\alpha^2)}$), the HS separability {\it probability} results,
\begin{equation} \label{HSsepprobcases}
\begin{cases}
 -\frac{1}{4} \alpha  (\alpha +1) & -1<\alpha <0 \\
 \frac{5}{8} \alpha  (\alpha +1) & \alpha +\sqrt{2}=0 \\
 -\frac{1}{8} \left(-5+\sqrt{5}\right) \alpha  (\alpha +1)
   & 2 \alpha +1=\sqrt{5} \\
 -\frac{1}{4} (\alpha -2) \alpha  (\alpha +1) & 0<\alpha
   <\frac{1}{2} \left(-1+\sqrt{5}\right)\lor \frac{1}{2}
   \left(-1+\sqrt{5}\right)<\alpha <1 \\
 \frac{3}{4}-\frac{1}{4 \alpha } & \alpha \geq 1 \\
 -\frac{1}{4 \alpha }+\frac{3}{4}+\frac{1}{\alpha -1} &
   \alpha +\sqrt{3}\leq 0 \\
 -\frac{(\alpha +1) \left(\alpha ^4-5 \alpha
   ^2+1\right)}{4 \alpha } & -\sqrt{2}<\alpha <-1\lor
   -\sqrt{3}<\alpha <-\sqrt{2}
\end{cases}
\end{equation}
In Fig.~\ref{fig:HSsepprob}, we plot these rather interesting/intricate 
results.
The separability
probabilities are {\it zero} at the isolated points $\alpha=-1,0$. 
For $\alpha \to  \pm  \infty$, a maximum of $\frac{3}{4}$ is approached. 
We also see that the ``golden ratio'' (or ``golden mean'') \cite{livio,markowsky} 
[or its inverse, depending upon
the definition], $\frac{\sqrt{5}-1}{2}$,  enters into delineating the 
different segments over which the separability probabilities  take different 
functional forms. (It would seem plausible, although we have not conducted
a full, detailed analysis that the points at which the functional forms
change, correspond to separability constraints
passing from inactive to active roles, and {\it vice versa}.)

{\it Fake} entanglement is avoided in this three-parameter 
model \cite[p. 128]{canosa} \cite[p. 4]{canosa2}. 
It did not appear feasible to directly expand the {\it trivariate} 
$(b_{q}, \sigma_{q}^2, \alpha)$ 
scenario just investigated to a {\it quadrivariate} one.
\subsubsection{Bivariate Jaynes state using generalized Bell-CHSH observables} \label{bivariatesec}
\begin{figure}
\includegraphics{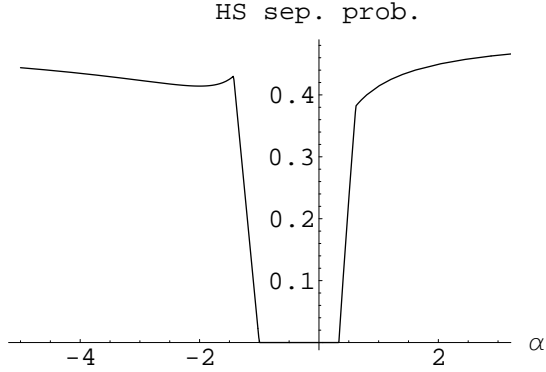}
\caption{\label{fig:HSsepprob2}Separability probabilities 
(\ref{anotherinstance}) based on the Hilbert-Schmidt metric, 
as a function of the parameter $\alpha$,
for the (Canosa-Rossignoli) class of {\it bivariate} Jaynes states employing
{\it generalized} Bell, Clauser, Horne, Shimony, Holt (Bell-CHSH) observables
(\ref{canosanew}). The variance $\sigma_{1}^2$ has been set to its value
$4 (1 +\frac{b_{1}^2}{8})$ in the single constraint (Horodecki) case}
\end{figure}

Let us, however, consider a related {\it bivariate} 
(Canosa-Rossignoli-type \cite[p. 126]{canosa}) model, in which we set
$\sigma_{1}^2 = 4 (1+\frac{b_{1}^2}{8})$, 
the dispersion in the single constraint
case \cite[eq. (13)]{Raj}. Then, the HS separability probabilities take the 
form (the HS volume here being 
$\frac{2 \sqrt{3 \alpha ^4-2 \alpha ^2+3}}{4 \alpha ^2+4 \alpha }$),
\begin{equation} \label{anotherinstance}
\begin{cases}
 -\alpha -1 & \frac{1}{3} \left(1-2 \sqrt{7}\right)<\alpha
   <-1 \\
 \sqrt{(\alpha -2) \alpha }-\sqrt{\alpha  (\alpha +1)}-1 &
   3 \alpha +2 \sqrt{7}\leq 1 \\
 \sqrt{\alpha  (\alpha +1)}-\alpha  & 2 \alpha +1\geq
   \sqrt{5} \\
 -\alpha +2 \sqrt{\alpha  (\alpha +1)}-1 &
   \frac{1}{3}<\alpha <\frac{1}{2}
   \left(-1+\sqrt{5}\right).
\end{cases}
\end{equation}
We represent this in Fig.~\ref{fig:HSsepprob2}.
In the $\pm \infty$ limit, the separability probability approaches $\frac{1}{2}$.
The HS separability probability is zero in the interval $\alpha
 \in [-1,\frac{1}{3}]$. The point $\alpha=1$ corresponds to the use
of the standard (ungeneralized) Bell-CHSH observable ($B_{1}$), and to
the {\it one}-constraint Horodecki model \cite{horodeckis}. We can see that
the associated HS separability probability is the silver mean.

\subsection{Tsallis-Lloyd-Baranger Two-Qubit States} \label{TLBsec}
Tsallis, Lloyd and Baranger have considered a scenario in which the probabilities of being in either one of the four states of the Bell basis is given
in the form $\frac{(1-x)}{4}, \frac{(1-y)}{4}, \frac{(1-z)}{4}$
and $\frac{(1+x+y+z)}{4}$ \cite{tlb}. (The feasible points $(x,y,z)$ lie in a 
certain tetrahedron \cite[Fig. 3]{tlb}.) 
They also embed their three-parameter
 (two-qubit) 
$4 \times 4$ density matrix \cite[eq. (12)]{tlb} into an (unnormalized)
four-parameter $4 \times 4$ density matrix \cite[eq. (14)]{tlb} 
by introducing the $q$-parameter. 

Upon its normalization and application of the H{\"u}bner formula (\ref{hub2}),
we obtained the corresponding Bures metric, the volume element of which,
in a numerical investigation, appeared to be zero, in this four-parameter
extended case.
So, we have, to this point, yet to find any {\it nondegenerate} $q$-extension 
of the Bures metric (if one is so possible).

In the unextended three-parameter case, if we employ new coordinates
of the form,
\begin{equation}
x=1 - 4 \cos{\theta_{1}}, y=1 - 4 \sin{\theta_{1}} \cos{\theta_{2}}, z= 1 - 4 
\sin{\theta_{1}} \sin{\theta_{2}} \cos{\theta_{3}},
\end{equation}
then we have simply
\begin{equation}
d_{TLB}(\rho,\rho+ d \rho)^2= d \theta_{1}^2 + \sin{\theta_{1}}^2
d \theta_{2}^2 + \sin{\theta_{1}}^2 \sin{\theta_{2}}^2 d \theta_{3}^2,
\end{equation}
that is, the {\it uniform} metric on the 3-sphere.

The Bures {\it volume} of the (separable and nonseparable) 
TLB-states is $\frac{\pi^2}{8}$, while the Bures volume of just 
the separable states themselves is (thanks to a challenging 
computation  --- involving a {\it cylindrical algebraic decomposition} [cad]  
\cite{cylindrical} --- performed by
M. Trott) $\frac{\pi (4 -\pi)}{8}$. 
(The separable states comprise the cube $x,y,z \in [-1,\frac{1}{3}]$.)
Thus, the Bures 
separability probability \cite{slaterEuro,slaterJGP,slaterPRA} 
of the TLB-states is (quite elegantly)
$\frac{4-\pi}{\pi} \approx 0.27324$.

For the Hilbert-Schmidt metric, the volume of the TLB (separable and
nonseparable) states is $\frac{1}{6 \sqrt{2}}$ and that of the separable states, $\frac{1}{12 \sqrt{2}}$, so the HS separability probability is $\frac{1}{2}$.

\section{Rajagopal-Abe Metric}

We also investigated the possible application of the ``generalized 'metric'' 
(based on the $q$-Kullback-Leibler entropy) \cite[eq. (16)]{RajAbe} 
to the two-level quantum systems (\ref{nonescortDensityMatrix}) --- and seeing how it pertains to the family of quantum 
monotone metrics \cite{petzsudar}.
For the case $q=1$ (which should reduce to the 
Kullback-Leibler symmetrized divergence \cite[eq. (6)]{RajAbe}, 
our calculations yielded that 
the diagonal elements of the ``metric'' take the form
\begin{equation}
\frac{1}{2} \Big(-\log{(-\frac{r}{2 +r})} +\log{W} \Big) dr^2 
-\frac{1}{4} r  (3 +\cos{2 \theta_{1}} +2 \sin{2 \theta_{1}}) \log{W}
d \theta_{1}^2 
\end{equation}
\begin{displaymath}
 -\frac{1}{4} r  (3 +\cos{2 \theta_{2}} +2 \sin{2 \theta_{2}}) 
\log{W} \sin^2{\theta_{1}}
d \theta_{1}^2,
\end{displaymath}
where, as we recall, $W=\frac{1-r}{1+r}$.
So, it clearly can not possess the form required of a quantum monotone metric 
(cf. Sec.~\ref{sec3D}).
However, when we attempted to implement equation (6) of (\cite{RajAbe}), bypassing the $q$-framework, we obtained for the (presumably same?) metric
\begin{equation}
\frac{1}{1-r^2} dr^2 -\frac{1}{2 r} \log{W} dn^2,
\end{equation}
which does not appear to correspond to a {\it monotone} metric.
\section{Concluding Remarks}
Naudts \cite{naudts} 
introduced the concept of a $\phi$-exponential family of 
density operators $\rho_{\theta}$ (for which the obvious example is $\phi(u)=u^q$).
He showed that the $\phi$-exponential family of density operators, together
with a family of escort density operators, optimizes a generalized version
of the well-known Cram\'er-Rao lower bound. He assumes that certain
Hamiltonians are two-by-two commuting. Therefore, the quantum information
manifold $(\rho_{\theta})_{\theta}$ is abelian, which ``is clearly too
restrictive for a fully quantum-mechanical theory''. He suggests further work to remove this restriction.

Abe regarded the order of the escort distribution $q$ as a parameter \cite{abe1}. He
studied the geometric structure of the one-parameter family of
escort distributions using the Kullback divergence, and showed that the
Fisher metric is given in terms of the generalized bit variance, which
measures fluctuations of the crowding index of a multifractal.

\begin{acknowledgments}
I wish to express gratitude to the Kavli Institute for Theoretical
Physics (KITP)
for computational support in this research and to Michael Trott 
of Wolfram Research Inc. for
his generous willingness/expertise in assisting with Mathematica computations.

\end{acknowledgments}

\bibliography{Canosa}

\end{document}